\documentclass[11pt]{article}
\usepackage{xspace}
\usepackage{graphicx}
\usepackage{amsmath}
\usepackage{amssymb}
\usepackage{color}
\usepackage{feynmp}
\usepackage{setspace}

\renewenvironment{thebibliography}[1]{%
   \begin{oldthebibliography}{#1}%
     \setlength{\itemsep}{-.95ex}%
}%
{%
   \end{oldthebibliography}%
}

\definecolor{Red}{rgb}{1,0,0}
\definecolor{Green}{rgb}{0,1,0}
\definecolor{Blue}{rgb}{0,0,1}
\definecolor{Black}{rgb}{0,0,0}

\def\beq{\begin{equation}}
\def\eeq#1{\label{#1}\end{equation}}
\def\eeqn{\end{equation}}

\def\beqa{\begin{eqnarray}}
\def\eeqa#1{\label{#1}\end{eqnarray}}
\def\eeqan{\end{eqnarray}}

\let\bar=\overbar

\def\Dslash{\not{\hbox{\kern-4pt $D$}}}
\def\dslash{\not{\hbox{\kern-2pt $\del$}}}

\def\msb{{\bar{\ssstyle M \kern -1pt S}}}

\def\Title#1{\begin{center} {\Large {\bf #1} } \end{center}}

\begin{document}

\Title{The Search for Magnetic Monopoles}

\bigskip\bigskip


\begin{raggedright}  

James L. Pinfold\index{Pinfold, J.}, {\it Physics Department, University of Alberta}\\

\end{raggedright}

{\small
\begin{flushleft}
\emph{To appear in the proceedings of the Interplay between Particle and Astroparticle Physics workshop, 18 -- 22 August, 2014, held at Queen Mary University of London, UK.  }
\end{flushleft}
}

\section{Introduction}

You could say that the idea of a magnetic monopole started in 1269. In that year, the scholar, soldier and
monk, Pierre de Maricourt produced a remarkable document named the Epistola de Magnete \cite{Epistola} that
identified for the first time that a magnet had a north and a south pole. This begs the age--old
question: can there be a single pole, a magnetic monopole? By the 19th century Gauss's law
 for magnetism, enshrined as one of Maxwell's
equations, stated mathematically that magnetic monopoles do not exist.
Maxwell's equations - that posit only electric charges occur in nature  - can
be made fully symmetric under the interchange of the electric and magnetic fields
if magnetic charges also exist. A courageous Pierre Curie was the first to suggest that
 magnetic monopoles could conceivably be present in nature in a paper published in 
 1894 \cite{Curie}.
 
  It was Dirac who took up Curie's challenge.
In 1931, he hypothesized that a magnetic monopole could exist within the framework of
quantum mechanics \cite{Dirac}. He saw the monopole as the end of an infinitely thin infinitely long solenoid
called a ``Dirac string''. The annoying string does not pose a problem as long as it
cannot be detected. The mathematics that ensured this gave rise to a quantum relation between
the electric and magnetic charges:   
$   e^{iqg} =1 \implies qg = 2\pi n (n=1,2,3...)$
where $q$ represents the charge of the electron and $g$ the Dirac (magnetic)  charge. This relation
is  the Dirac Quantization Condition that expresses  the quantization of electric charge, as long as
at least one monopole exists.  In 1969 Schwinger published a paper \cite{Schwinger} describing an extension of 
  the Dirac quantization condition  to a dyon - a hypothetical particle that has both magnetic and 
  electric charge.

In 1974 Gerard Ôt Hooft \cite{thooft}  and Alexander Polyakov \cite{Polyakov} 
discovered that model for Electroweak Unification proposed by Georgi and Glashow \cite{Georgi} 
contained a stable  field configuration - like a knot  that cannot unravel - corresponding to a magnetic 
monopole. The 't Hooft-Polyakov monopole is  totally different from the Dirac monopole,  it is 
a topological defect or a topological soliton.   The mass of such a  monopole - roughly 100 GeV,  determined 
  by the energy scale associated with weak nuclear forces - is too low to be compatible with experimental results. 
  But 't Hooft and Polyakov noticed that the structure of the SU(5) Grand Unified Theory (GUT) was similar to that 
  of the Georgi-Glashow model and hence also contained a monopole solution, but with the much higher mass of 
  10$^{16}$ GeV/c$^{2}$.   The 't Hooft-Polyakov, or GUT,  monopoles also have the exciting  ability  to catalyse 
nucleon decay \cite{pdecay}. The heart of the monopole retains the full SU(5) symmetry that held
prior to the freezeout of the electroweak and strong forces, when leptons and quarks were part of the same
extended family. This means that when a proton or a neutron gets in contact with a GUT monopole, 
it is very likely to decay. 
  
  The discovery of a Higgs-like boson in July 2012 by the ATLAS and CMS \cite{Higgs}  experiments at the
Large Hadron Collider (LHC) has lead unambiguously to the completion of the puzzle of the
Standard Model spectrum.      It has been asserted that the Weinberg-Salam model has no topological 
  monopole of physical interest.  But, Cho and Maison \cite{Cho}  pointed out that the Weinberg-Salam model 
  could be viewed as a gauged CP$^{1}$ model, with the Higgs doublet  field interpreted as the CP$^{1}$ field. 
  This would be an extension of our view of the Standard Model. Importantly, the second homotopy group 
  of the gauged CP$^{1}$ model is the   same as that of the Georgi-Glashow model that contains
   the 't Hooft-Polyakov monopole.

 The Cho-Maison monopole is the electroweak (EW) generalization of 
Dirac's monopole, that can be viewed as a hybrid of Dirac and Ôt Hooft-Polyakov monopoles.
But unlike  DiracÕs monopole, it carries the magnetic charge 4$\pi$/e. This is because in the Standard
Model the U(1)$_{EM}$ part  has the period of 4$\pi$ not 2$\pi$, as it comes from the U(1) subgroup of SU(2). Thus
the magnetic charge of the EW monopole is twice that of the Dirac Monopole. Recent estimates of
the EW monopole mass  \cite{Cho-mass} indicate that it's  possibly  detectable at the LHC.
  
Since 1931 physicists have been assiduously searching for magnetic monopoles: in cosmic
rays; trapped in bulk matter, in lunar dust, meteors and on Earth; also, at accelerators
where they would be produced in high-energy particle interactions. Indeed, a monopole
search has been performed at each advance of the high-energy frontier driven by
the relentless increase of particle accelerator energy. MoEDAL -  the Monopole and
Exotics Detector at the Large Hadron Collider (LHC) -  has been custom-made
to continue the search, along with the general-purpose LHC detectors, to the
multi-TeV realm.

\section{Monopole Properties}

Dirac showed the magnetic charge of a monopole is $g = n(e/2\alpha$), where $\alpha$ ($\approx$ 1/137) is the
 fine structure constant.  Thus, a unit of magnetic monopole is $g \sim$ 68.5e. The magnetic charge should be larger 
 if n$ > $ 1 and also if the fundamental  unit of electric  charge is $e/3$, rather than $ e$. In analogy with the fine
  structure constant, $\alpha = e^{2}/\hbar c$, the dimensionless magnetic coupling constant is
   $\alpha_{g} = g^{2}_{D}/\hbar c  = $34.25  which is much larger than one and thus perturbative 
   calculations  involving monopoles face serious difficulties. 

As we can see from the Bethe formula for magnetic charge given below 
the energy loss of a singly charged (n=1) very relativistic magnetic monopole  ($\beta \approx$ 1) is amazingly about 4700 times 
that of a proton! The Bethe formula for ionization energy loss by a relativistic monopoles  is given by replacing $ze$ with $ng\beta$:
\begin{equation}
-\frac{dE}{dx} = (ng/e)^{2}\frac{KZ}{A}\left[\frac{1}{2}ln\left(\frac{2m_{e}\beta^{2}\gamma^{2}}{I^{2}} \right) - \beta^{2} \right]
\end{equation} 
 Also, the energy loss for magnetically charged particles falls with decreasing $\beta$ unlike electrically charged 
particles where the energy loss rises quickly  when  $\beta\gamma \lesssim$1.3.

A moving magnetic monopole  traversing the superconducting wire coil of a SQUID (Superconducting Quantum Interference 
Device)   drives an electrical current within the coil. One can use Faraday's Law to calculate the magnitude of the current that 
would be  generated: $I = -\mu_{0}g/L$, where L is the inductance of the coil and $\mu_{0}$ is the permeability of free space. 
Consequently, the  induced currenr only   depends on the magnetic charge and is independent of the speed or direction of the 
magnetic monopole. 
  

Monopoles will accelerate along magnetic field lines acquiring energy. The energy W acquired in a magnetic field B
is  $W = ng_D B l =$ n20. 5 GeV/T m, where $l$ is the distance travelled in a coherent magnetic field. 
Thus, modern magnets can be easily be used to accelerate monopoles  to multi-TeV energies. Another unique signature 
of a monopole is its usual trajectory in a magnetic field. For example in a solenoidal field  electrically charged
particles curve in a plane transverse to the field lines (the $r-\phi$ plane)  but do not bend in the plane running
parallel with the field lines (the $r-Z$ plane). Conversely,  magnetic monopole move along parabolic paths in the $r-z$ plane but
no not bend in the $r-\phi$ plane.

\begin{figure}[!ht]
\begin{center}
\includegraphics[width=7.5cm]{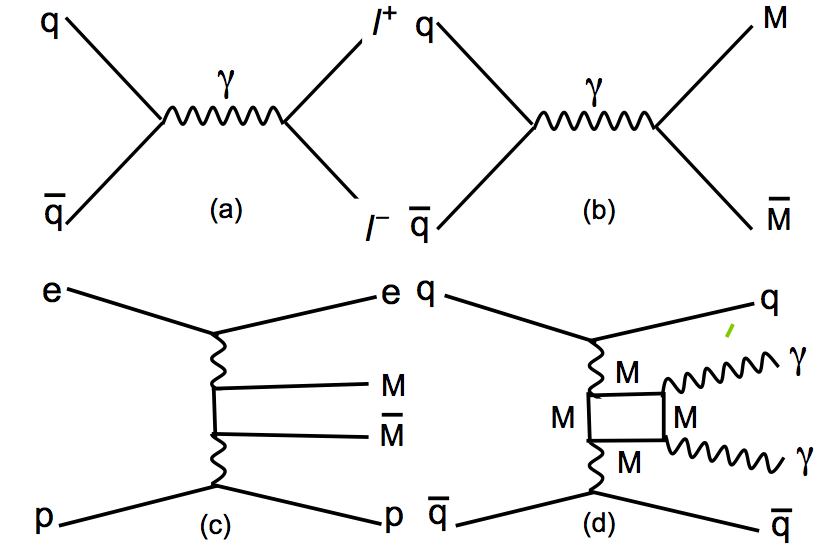}
\caption{(Left) (a) Drell-Yan production of a lepton pair. (b) Drell-Yan production of a monopole pair. (c) Two photon 
productionof a monopole pair. (d) Two photon production of a high energy photon pair production via a virtual
 monopole box.}
\label{fig:feynman}
\end{center}
\end{figure}

\section{Searching for  Magnetic Monopoles at Colliders}
Magnetic  Monopole searches have been performed  at colliders \cite{14},
 in cosmic rays and in matter since the 1950s, and are still 
continuing today  \cite{ATLAS}.  Table~\ref{acc-searches} gives a list
of the accelerator searches at accelerators over the last 25 years. Most searches
at accelerators rely on the highly ionizing nature of the magnetic monopole, utilizing 
the following detector technologies: scintillators, gaseous counters and Nuclear Track Detectors (NTDs).

 Due to the large coupling constant of the magnetic monopole perturbative calculations of the magnetic
monopole cross section cannot be made. Instead a Drell-Yan mechanism is assumed for
the cross-section calculation. Figure~\ref{feynman}  shows a Feynman diagram of the Drell-Yan 
mechanism for dimuons and monopole-antimonopole production. These two diagrams shows
annihilation of the quark-antiquark via the intermediate virtual photon and later photon
decay into the two leptons (a)  and monopole-antimonopole pair (b). Monopoles pairs
can also be produced in two photon interactions, as shown  in Figure~\ref{feynman}(c).

The search for Monopole production in nucleus-nucleus (AA) collisions 
 has been proposed \cite{ions} and performed \cite{ion-search}.
This interest is motivated by the expected $\sim Z^{4}$ enhancement 
in  production \cite{z4} of magnetic monopoles. But, this benefit of this enhancement
 is countered by the rapid fall in the intensity of   interaction with the produced particle mass \cite{127}.

A virtual process  involving a monopole box diagram \cite{Ginzburg} that gives rise to two
high transverse momentum photons - as shown in Figure~\label{feynman}(d) - was sought by the D0
 collaboration \cite{D0}. Such a process can be produced in  $e^{+}e^{-}$ or $q\overline{q}$  collisions.
 The L3 experiment at LEP \cite{L3} searched for Z decays into 3 photons that can take place via a monopole
 loop.  However, calculations involving monopole loops are challenging due to the size of the monopole's 
 magnetic charge and coupling constant. Consequently,  indirect limits based
 on this approach are questionable \cite{Problems}.

\begin{table}[!th]
\begin{center}
\caption{A list of  monopole searches at accelerators over the last 25 years.}
\scalebox{0.8}{
\begin{tabular}{| l |  l | l | l | l | l | l | l | }  \hline\hline
  Reaction &    $\sqrt{s}$ & M$_{limit}$   & $\sigma$ MM  & MM       & Tech.     & Year & Ref. \\ 
                   &  GeV             & GeV/c$^{2}$      &    cm$^{2}$     &   Chrg      &               &            & \\  \hline
pp             &        1800      &  $ <  $850              & 2.e-34           & $\ge$0.5    & Plstc     & 1990  & \cite{23} \\ \hline
e+e-         &        88-94     &  $ <  $45                & 3.e-37           &  1                 & Plstc     &  1992 & \cite{32} \\ \hline
e+e-         &        88-94    &                             &                       &                     & Plstc     & 1993  & \cite{33} \\ \hline
pbA          &        17.9       &  $ < $ 8.1                & 1.9e-33        &  $\ge$ 2      & Plstc     & 1997  & \cite{18d} \\ \hline
AuAu       &          4.87     &  $  <  $3.3                 & 0.65e-33     &  $\ge$ 2      & Plstc    & 1997   & \cite{18d} \\ \hline
ppbar      &       1800       &  260-420           & 7.8e-36         &  2-6             & Indctn  & 2000  & \cite{19}  \\ \hline
ppbar      &       1800      &   265-410           & 0.2e-36         & 1-6              &  Indctn & 2004  & \cite{20} \\ \hline
e+p          &         300       &                            &  0.5e-37        &  1-6             & Indctn  & 2005  & \cite{22} \\ \hline
ppbar      &      1800       &    369                  & 0.2e-36         & $\ge$1       & Cntr      & 2006  & \cite{34} \\ \hline
e+e-         &       206.3    &    45 - 102          & 0.05e-36      &     1               & Cntr     & 2007  & \cite{LEP2} \\ \hline
pp            &       7000     &     200-1500       & 2e-39            &     1               & Cntr     & 2012  & \cite{ATLAS} \\ \hline
\end{tabular}%
}
\label{acc-searches}
\end{center}
\end{table}

\section{The Search for Monopoles for  Cosmic Monopoles}
Cosmic monopoles that are detectable on Earth  can either be light, for example produced locally  in
cosmic ray interactions  with the atmosphere  with mass less significantly less 
that the GUT scale, or else so heavy that they can only have been produced
 soon after the birth of the Universe in the Big-Bang. Most 
 of these searches have been based on the premise that the  monopoles  are produced
 in a  symmetry-breaking phase transition in the early Universe  as topological defects via the 
 Kibble mechanism \cite{kibble} -  typically with the GUT mass $\sim$,  10$^{16}$ GeV/c$^{2}$.
 Some GUT models and some supersymmetric models predict Intermediate Mass Monopoles (IMMs)
with masses 10$^{5}  <$ MM$_{mass}$  $<$ 10$^{12}$ GeV and with magnetic charges of multiples
 of $g_{D}$; these IMMs may  have been produced in later phase transitions in the early Universe and 
 could be present in the cosmic radiation \cite{lightGUT}.

\begin{table}[!th]
\begin{center}
\caption{A list of cosmic monopole  searches performed over the last 25 years.}
\scalebox{0.8}{
\begin{tabular}{| p{2cm} | p{2.5cm} | c | p{1.8cm} | c | c |}  \hline\hline
 Lab.    &    $\phi$(cm$^{-2}$sr$^{-1}$s$^{-1}$)  &    Comments            &  Technique       &   Year     & Ref. \\  \hline
             &   $<$ 7.2e-13                               &   all $\beta$                                     & Induction                     & 1990  & \cite{z1} \\ \hline
             & $< $ 1.e-18                                  &   3x10$^{-4} < \beta < $1.5x10$^{-3}$ & Mica               & 1990  & \cite{a2} \\ \hline
             & $<$ 1.8e-14                                & $\beta >$ 1.1 x 10$^{-4}$              & He-PT   & 1990 & \cite{b2} \\ \hline
Baikal  &   $ <$ 5.e-16                                &    $\beta < $ 10$^{-3}$                   & Catalysis & 1990 & \cite{c2} \\ \hline
           &   $<$ 3.8e-13                                &   all $\beta$                                     & Induction            &   1991           & \cite{d2} \\ \hline
            &  $ <$ 3.2e-16                                &  $\beta > $0.05                             & Plastic & 1991 & \cite{e2} \\ \hline
            & $<$ 7.2e-13                                  &      all $\beta$                                  & Induction                   & 1991  & \cite{f2}  \\ \hline
            & $<$ 4.4e-12                                  &      all $\beta$                                  & Induction                   & 1991  & \cite{g2}  \\ \hline
SOUDAN-2 &   $<$ 8.7e-15                       &  $\beta >$  2 x 10$^{-3}$                &  Counters                   & 1992  & \cite{h2} \\ \hline
 IMB     &  $<$ 8.7e-15                                &  $\beta \sim$ 10$^{-3}$                 & Catalysis &  1994 & \cite{i2} \\ \hline
MACRO & $<$ 5.6e-15                              & $\beta$ = (1.8 - 3.0) x 10$^{-3}$  & Plstc/cntrs & 1994 & \cite{j2} \\ \hline
MACRO & $<$ 1.e-15                                & $\beta$ = 0.18 - 3.0x10$^{-3}$   & Plstc/cntrs & 1997 & \cite{k2} \\ \hline
MACRO & $<$ 1.5e-15                              &  5x10$^{-3} < \beta <$  0.99      & Plstc/cntrs & 2002 & \cite{l2} \\ \hline
MACRO & $<$ 1.4e-16                              &   1.1x10$^{-4} < \beta <$ 1         &  Plstc/cntrs & 2002 & \cite{m2} \\ \hline
MACRO & $<$ 3.e-16                                 &   1.1x10$^{-4} < \beta <$ 5x10$^{-3}$ & Catalysis  & 2002   & \cite{n2} \\ \hline
RICE &$<$1.e-18                                        &  $\gamma >$ 10$^{8}$                  & Cerenkov      &    2008        & \cite{o2} \\ \hline
SLIM & $<$ 1.3e-15                                   & $\beta >$0.05                                  & Plastic     & 2008             & \cite{p2} \\ \hline
AMANDA2 & $<$ 3.8e-17                        &  $\beta >$0.76                                 & Cerenkov             &    2010   & \cite{q2} \\ \hline
ANITA2  & $<$ 1.e-19                               &  $\gamma >$ 10$^{10}$                & Cerenkov             &  2011   & \cite{r2} \\ \hline
Super-K &  $<$6.e-28                               &       $\beta =$ 10$^{-5}$                 & Catalysis & 2012 & \cite{s2} \\ \hline 
ANTARES  &$<$1.3e-17                              &  $\beta >$ 0.625                              & Cerenkov                    & 2012  & \cite{t2} \\ \hline
IceCube & $ <$ 3.e-18                               &    $\beta >$ 0.8                                  &  Cerenkov                  &   2013 & \cite{u2} \\ \hline
IceCube &  $<$  1.e-18                              &                                                              & Catalysis & 2014 & \cite{v2} \\ \hline                                     
\end{tabular}
\label{nonacc-searches}
}
\end{center}
\end{table}

A wide variety of techniques have been   employed to search for cosmic
monopoles,  these include: induction techniques with SQUID magnetometers to detect
magnetic charge trapped in matter or in flight; detector arrays based on ionisation energy loss;
 detection of Cherenkov radiation in water or ice to probe relativistic monopoles;  and nucleon-decay 
 detectors for probing the monopole catalysis reaction.  A list of experiments designed to search
 for cosmic ray monopoles over the last 25 years is given in Table~\ref{nonacc-searches}.
 
Cosmic monopoles will be accelerated by galactic and intergalactic magnetic fields, draining energy
from these fields.   Parker \cite{parker}  obtained an upper bound on the flux of monopoles in the galaxy 
based on the contemning existence of these fields. With reasonable choices for the astrophysical parameters
 \cite{reasonable} ,  this Parker bound is: $F < 10^{15} cm^{-2}sr^{-1}s^{-1}$, when the monopole
 mass is less than around  10$^{17}$ GeV.
 
 \subsection{Observations of   Cosmic Monopole Candidate Events  Using SQUIDs}
 
In the early 1980Õs,   Blas Cabrera was one of the first experimenters to deploy a SQUID device 
 in order to detect magnetic monopoles from the  cosmos. His initial detector employed a four-turn, 
 5-cm-diam loop, positioned with its axis vertical,  connected to the superconducting  input coil of a SQUID magnetometer.
During the night of February 14, 1982, this detector recorded an event that was completely consistent with being due to a
magnetic monopole \cite{cabrera}. A  group at Imperial College London observed another 
perfect candidate magnetic monopole   event on the 11th of August 1985 \cite{caplin}.
To this day it is still not clear what background process can explain these signals.

\section{The MoEDAL Experiment at the LHC} 
MoEDAL, the 7th and newest experiment \cite{TDR} is dedicated to the detection of the highly ionizing
particle avatars of new physics such as the magnetic monopole and massive stable or metastable
charged particles. Such particles originate from a number of Beyond the Standard Model (BSM) scenarios
that, for example, incorporate: magnetic charge, new symmetries of nature (eg Supersymmetry); extra
spatial dimensions; dark matter particles, etc. The MoEDAL detector is a largely passive detector that
makes it totally unlike other collider detectors. Essentially MoEDAL is a largely passive detector, like a 
giant camera ready to reveal ``photographic''   evidence  for new physics - where plastic Nuclear Ttrack 
Detectors are the film.  MoEDAL also has trapping  detector volumes capable of capturing long-lived 
electrically and magnetically charged particles from beyond the  Standard Model for further 
 monitoring and study at a remote SQUID magnetometer 
facility and a deep  underground laboratory  such as SNOLAB. The MoEDAL experiment will significantly 
expand the horizon for discovery of the  LHC, in a complementary way.  It will start to
 take data in the  Spring of 2015  when the LHC restarts at the unprecedented energy of  $\sim$ 14 TeV
 
 \begin{figure}[!ht]
\begin{center}
\includegraphics[width=10.5cm]{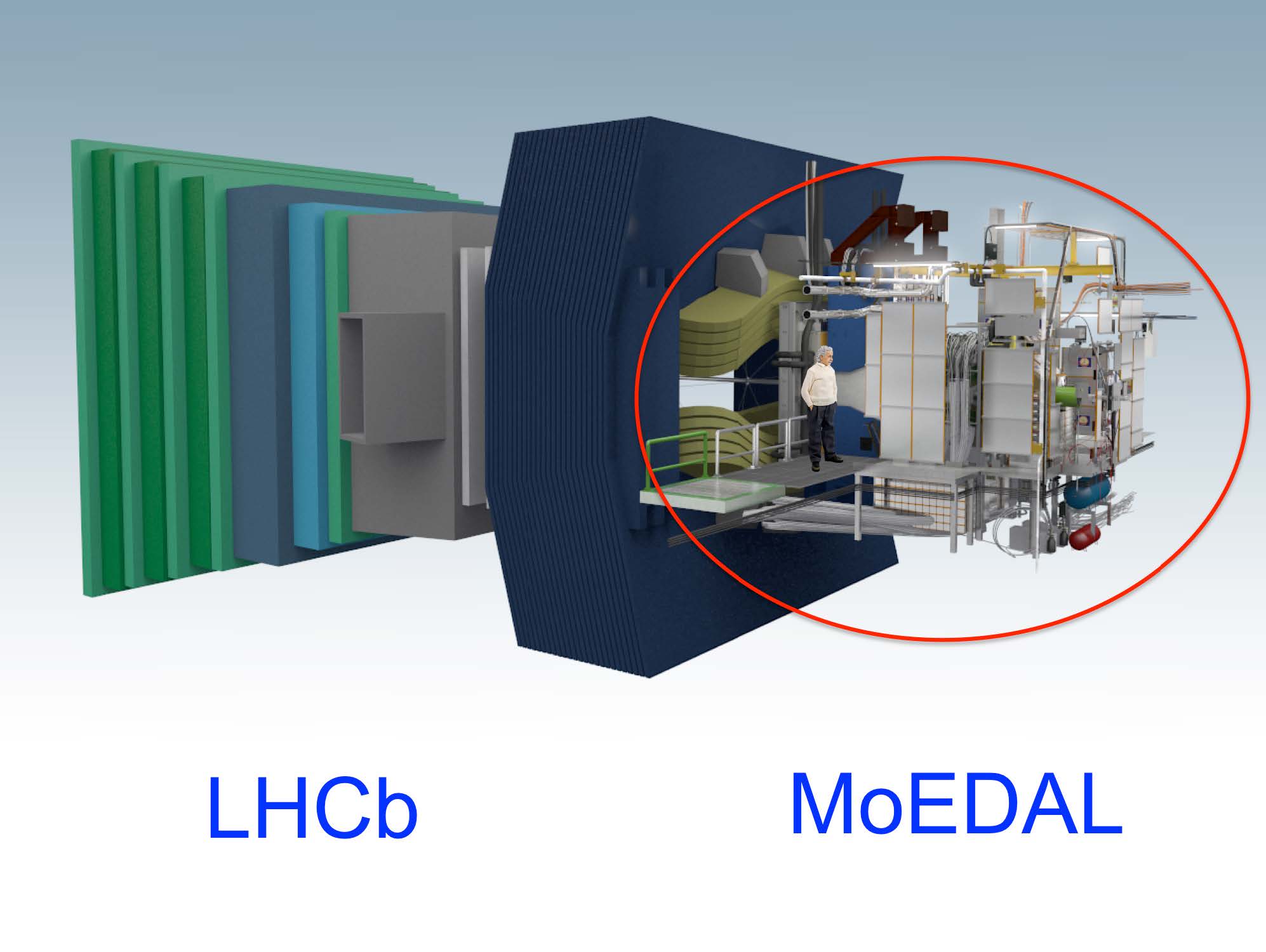}
\caption{(Left) The MoEDAL experiment surrounding interaction sharing the relatively open
 interaction point 8 on the LHC ring with the LHCb experiment.}
\label{fig:moedal}
\end{center}
\end{figure}

The main LHC experiments are optimized to detect conventionally charged relativistic particles produced with a velocity
 high enough for them to be fully registered in the detector in  the time between  bunch crossings. Exotic, highly ionizing 
 detector-stable  particles  produced at the LHC might be sufficiently massive to 
 travel through the detector  within a single  trigger window and so will have a low   efficiency for detection. Also, 
 the sampling time and reconstruction   software of each sub-detector is optimized  assuming that particles are
  travelling at close to the velocity of light.   Hence, the quality of the read-out signal,  reconstructed 
  track or cluster may be degraded for an SMP, especially    for subsystems at some distance from the
   interaction point. Another problem is that the main LHC detectors cannot
   be directly calibrated for highly ionizing  particles such as the magnetic monopoles.

An additional  challenge is that very highly ionizing particles can be absorbed before they penetrate the detector fully. 
Additionally, the read-out electronics of conventional LHC detector systems are usually not designed to have a
 wide enough dynamic range to measure the very large dE/dx of highly ionizing particles properly. In the case
  of the magnetic monopole there is also the problem of understanding the response of conventional LHC detector 
  systems to particles with magnetic charge. All the above drawbacks can only get worse for increasing magnetic charge.
  
\begin{figure}[!ht]
\begin{center}
\includegraphics[width=8.5cm]{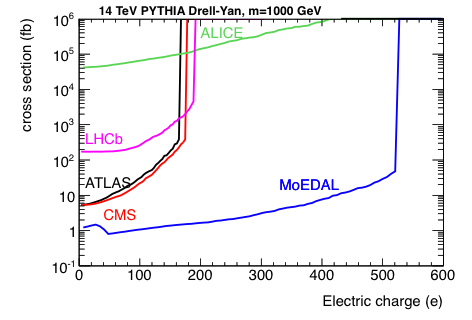}
\caption{The expected reach  of the search for direct monopole - anti-monopole Drell-Yan pair production process at the 
LHC (Ecm =14 TeV). Assuming the luminosity taken by  LHCb/MoEDAL is 2 fb$^{-1}$, by  ATLAS \& CMS 20 fb$^{-1}$, and  
by ALICE 0.004 fb$^{-1}$.}
\label{fig:sensitivity}
\end{center}
\end{figure}

  The MoEDAL experiment bypasses these experimental challenges by using a passive plastic NTD technique
   that does not require a trigger. Also, track-etch detectors provide a tried-and-tested method to detect and measure 
   accurately the track of a very highly ionizing particle and its effective Z/$\beta$. Importantly, heavy-ion beams provide 
   a demonstrated calibration technique because they leave energy depositions very similar to those of the hypothetical 
   particles sought. A magnetic monopole would leave a characteristic set of 20 collinear etch-pits. 
   Uniquely,  a trapping detector array incorporated into MoEDAL allows the possibility of the direct detection of magnetic charge
   and the possibility to directly study the monopole.  In this case  monopoles   are  ranged out,  stopped and then captured in matter 
   by their interaction with the magnetic moment of  the detector nuclei,  forming  long-lived bound states \cite{trapping}
   There are no Standard Model  particles that could produce the distinctive signatures described above  Ð thus, even one
    event would herald   a discovery. A plot showing the
      estimated sensitivity of MoEDAL for magnetic monopoles compared to the other LHC detectors is shown 
      in Figure~\ref{fig:sensitivity}
    
  MoEDAL's groundbreaking and complementary physics program is presented in a recently published paper \cite{physics}. 
  This program covers over thirty fundamentally important Beyond the Standard Model (BSM) scenarios that place MoEDAL
  firmly at the Terascale  discovery frontier. A key physics aim of MoEDAL is the search for the magnetic Monopole/Dyon. 
   In addition, several    other magnetically charged messengers of new physics are described in the MoEDAL physics
    paper \cite{physics} and in    references therein.   Of major importance is MoEDALÕs sensitivity to a wide range of 
    massive detector stable, single,  fractionally (charge $>$1), or multiply electrically charged HIPs that arise from a 
    number of well motivated beyond the  Standard Model  scenarios.   More than 20 such topics are 
    described in MoEDALÕs physics program \cite{physics}.  These BSM arenas, include:   new symmetries of nature
     such as supersymmetry and left-right symmetry: extra spatial 
     dimensions; a fourth generation;  technicolour; vector-like confinement; long lived heavy quarks, non-commutative
      geometry; and multi-particle excitations such as Q-balls, strangelets, and quirks, etc.

\section{``Cosmic-MoEDAL'' a Proposal}
SLIM was the first experiment to extend the search for cosmic   Monopoles with masses well below  the GUT scale, 
with a high sensitivity. However, SLIM's modest  size ($\sim$ 400 m$^{2}$)
  precluded  it searching for a flux of cosmic monopoles below the Parker Bound. In this conference and at other
  venues I proposed a very large array ($\gtrsim$ 10,000 m$^{2}$) of CR-39 detectors -`` Cosmic-MoEDAL'' - to be deployed at very 
  high altitude, for example at  the  Mt Chacaltaya lab. in Bolivia with an elevation of 5,400 m. Such,
  an array would be able to take the search for  cosmic monopoles  with velocities $\beta \gtrsim$ 0.1, from the LHC's TEV 
  scale all the way to the GUT scale,  for monopole fluxes  well below the Parker Bound. 

 Large scale underwater and under ice neutrino telescopes (Amanda, IceCube, ANTARES, NEMO)  can also 
search for fast IMMs with $\beta \gtrsim $ 0.5 to below  the Parker bound \cite{wissing}. 
 However, searches for lighter  IMMs by Earth based detectors are essentially limited to downgoing particles \cite{derkaoui} 
 where these detectors have  to discriminate against the large background of cosmic ray muons.  Detectors such as 
 IceCube should be able to search for sub-relativistic  velocities when the GUT monopole can catalyze
nucleon decays along their trajectories (Rubakov-Callan effect). This effect depends on the gauge group of the respective
 GUT theory \cite{RC-gauge} and on other assumptions,  for example  on the fermion masses or the relative velocity between
  the quarks and the   monopole utilized \cite{rel-vel}. Importantly, this process is not possible for IMMs  with
  masses below 10$^{13}$ GeV \cite{noRC}. Cosmic-MoEDAL does not suffer from these  limitations.

\section{Closing Remarks} 
Although, the experimental evidence for the monopole is still lacking the theoretical requirement
for their existence has increased. Dirac's monopole was ÒniceÓ but not required. But  many modern
theories including Grand Unified Theories, String Theory, M - theory all require magnetic monopoles.
One of the world's leading string theorists, Joseph Polchinski, has generalized Dirac's connection between 
magnetic monopoles and charge quantization. He has posited that in any theoretical framework that requires
 charge to be quantized, there will exist magnetic monopoles. He also maintains that in any fully unified theory,
  for every gauge field there will exist electric and magnetic sources. Speaking at the Dirac Centennial 
  Symposium  in 2002, he commented that "the existence of magnetic monopoles seems 
  like one of the safest bets that one can make about physics not yet seen" \cite{polchinski}.

\end{document}